# Dynamics of two-level atom interaction with single-mode field

Sitotaw Eshete [a,*], Yimenu Yeshiwas [b]

[a] Physics Department, Debre Tabor University, Debre Tabor, Ethiopia,
[b] Physics Department, Debark University, Debark Ethiopia

**Abstract:** In this paper, the dynamics of fluorescent light emitted by a two-level atom interacts with squeezed vacuum reservoir is studied wisely using two-time correlation function fundamentals. The mathematical analysis shows the fluorescent spectrum of light emitted by the atom is turned out to be a single peak at a Lorentz's frequency. On the other hand, the squeezed vacuum reservoir input is responsible to the stimulated emission of photon from the atom. Moreover, it is identified that thermal reservoir is more efficient than squeezed vacuum reservoir to have valuable power spectrum.

## 1. Introduction

The Different quantum optical investigations have been made by various authors for several years. The power spectrum of a fluorescent light emitted by a two-level has been studied by enormous authors [1-8]. In connection with this, there is an author who investigates the dynamics of the power spectrum of the light scattered by a two-level atom [8]. The investigation clearly shows that the power spectrum of the scattered field is directly obtained from the dipole moment correlation function. In addition, Eyob and Fesseha considered that two-level atom interaction with squeezed light. According to their work, the fluorescent light is in squeezed state and the power spectrum consists of a single peak only [5].

Another examination on the interaction of a two-level atom with a single mode of radiation has been made by Cirac. Accordingly, with the aid of the master equation for the atomic density operator in a bad-cavity limit the light emitted by the atom into the background mode has been studied [7]. The interaction a system in which initially unexcited two-level atom with a weak cavity field has been another area regarding with two-level atomic interaction dynamic with single light mode [10].

Taking this as motivation, in this paper it is necessary to see how would be the dynamics of a fluorescent light emitted by a two-level atom but coupled to squeezed vacuum reservoir. Basic fundamental effects of squeezed vacuum reservoir are clearly stated in this paper. Here we derive the time evolution for atomic variables rather than light mode variable using the pertinent master equation. It contributes a very valuable scientific merit for the development of instrument empowerment which uses two-level atom interaction with single mode radiation principles.

## 2. Operator dynamics

Consider a single two-level atom present in an open cavity with squeezed vacuum reservoir. We denote the upper and lower electronic levels by $|a\rangle$ and $|b\rangle$, respectively. Applying the rotating wave approximation, the Hamiltonian of the cavity mode with the two-level atom is presented as

$$\hat{H} = i\lambda\left(\hat{\phi}_+\hat{a} - \hat{a}^\dagger\hat{\phi}_-\right) \qquad (1)$$

---

[*] Correspondence of the paper to: sitotaweshete11@gmail.com

with $\lambda$ is the atom with the cavity mode coupling constant. Besides, $\hat{\phi}_+ = |a\rangle\langle b|$ and $\hat{\phi}_- = |b\rangle\langle a|$ are the raising and lowering atomic operators, respectively. Moreover, $\hat{a}$ and $\hat{a}^\dagger$ are the photonic annihilation and creation operators for the cavity modes. The atomic operators are satisfying the commutation relations

$$[\hat{\phi}_+, \hat{\phi}_-] = \hat{\phi}_z, \tag{2}$$

$$[\hat{\phi}_\pm, \hat{\phi}_z] = \mp\hat{\phi}_\pm \tag{3}$$

and the operator $\hat{\phi}_z$ can be expressed in terms of the rising and lowering operators as

$$\hat{\phi}_z = \hat{\phi}_+ - \hat{\phi}_-. \tag{4}$$

It is very important to an expression for probability of the atom to be in the upper or lower electronic energy state. To this end, let's introduce the probability amplitudes denoted by $\rho_a$ (the probability of an atom to be in the upper electronic energy level) and $\rho_b$ (the probability an atom to be in the lower electronic energy level) which can be expressed as

$$\rho_a = \langle \hat{\phi}_+ \hat{\phi}_- \rangle = \frac{1}{2}\left(\langle \hat{\phi}_z \rangle + 1\right) \tag{5}$$

and

$$\rho_a = \langle \hat{\phi}_- \hat{\phi}_+ \rangle = \frac{1}{2}\left(1 - \langle \hat{\phi}_z \rangle\right). \tag{6}$$

At a time the atom can occupy whether the upper energy state or the lower energy state. Here we use many energy state approximation into only two-levels in which the atomic transition is resonant with the cavity mode, the sum of the probabilities given in Equations (5) and (6) turn out to be

$$\rho_a + \rho_b = 1. \tag{7}$$

This indicates that the sum of the probabilities of an atom yields one. This condition is agreed with the fact of probability theory. To study the dynamics of the considered system, we have developed the master equation which encompasses all the property of it. By taking consideration of the system interaction with the squeezed vacuum reservoir, we get the formal mathematical expression to the system as

$$\begin{aligned}\frac{d}{dt}\hat{\rho}(t) &= \frac{\gamma}{2}(N+1)\left(2\hat{\phi}_-\hat{\rho}\hat{\phi}_+ - \hat{\phi}_+\hat{\phi}_-\hat{\rho} - \hat{\rho}\hat{\phi}_+\hat{\phi}_-\right) \\ &+ \frac{\gamma}{2}N\left(2\hat{\phi}_+\hat{\rho}\hat{\phi}_- - \hat{\phi}_-\hat{\phi}_+\hat{\rho} - \hat{\rho}\hat{\phi}_-\hat{\phi}_+\right) + \gamma M\left(\hat{\phi}_-\hat{\rho}\hat{\phi}_- - \hat{\phi}_+\hat{\rho}\hat{\phi}_+\right)\end{aligned}, \tag{8}$$

where $\gamma$ is the atomic decay constant and the effects of the squeezed vacuum reservoir incorporated through the symbols $N$ (the reservoir mean photon number of the reservoir) and $M$.

## 3. Power Spectrum of a single-mode field

We now define the power spectrum of fluorescent light emitted by a two-level atom in an open space by

$$P(\omega) = \int_{-\infty}^{\infty} d\tau e^{i(\omega-\omega_0)\tau} \langle \hat{\phi}_+(t)\hat{\phi}_-(t+\tau) \rangle_{ss}. \tag{9}$$

In which $\omega_0$ is the central transition frequency of the atom from the upper level to the lower energy state and $ss$ stands for steady state conditions. It proves convenient that Equation (9) can be rewritten as

$$P(\omega) = \int_{-\infty}^{0} d\tau e^{i(\omega-\omega_0)\tau} \langle \hat{\phi}_+(t)\hat{\phi}_-(t+\tau) \rangle_{ss} + \int_{0}^{\infty} d\tau e^{i(\omega-\omega_0)\tau} \langle \hat{\phi}_+(t)\hat{\phi}_-(t+\tau) \rangle_{ss} \tag{10}$$

It is better to understand the two-time correlation function is not depending on the time $t$ rather it is depending only on the time difference $\tau$. Then let's replace $t$ by $t-\tau$ in the first integral of Equation (10) and performing the change of variables to get the power spectrum in the form

$$P(\omega) = 2\operatorname{Re}\int_0^\infty d\tau e^{i(\omega-\omega_0)\tau}\left\langle \hat{\phi}_+(t)\hat{\phi}_-(t+\tau)\right\rangle_{ss}, \tag{11}$$

where Re denotes the real part of the integral.

We next seek to determine the two-time correlation function involved in Equation (11) at steady state. To do this, we fist write the time evolution of atomic operators by employing the master equation given in Equation (8) and with the aid of trace properties, we have generated the following two Equations.

$$\frac{d}{dt}\left\langle \hat{\phi}_+(t)\right\rangle = -\frac{\gamma}{2}(2N+1)\left\langle \hat{\phi}_+(t)\right\rangle + \gamma M\left\langle \hat{\phi}_-(t)\right\rangle \tag{12}$$

and

$$\frac{d}{dt}\left\langle \hat{\phi}_-(t)\right\rangle = -\frac{\gamma}{2}(2N+1)\left\langle \hat{\phi}_-(t)\right\rangle + \gamma M\left\langle \hat{\phi}_+(t)\right\rangle. \tag{13}$$

According to Equation (4), we can express equations (12) and (13) as

$$\frac{d}{dt}\left\langle \hat{\phi}_+(t)\right\rangle = -\frac{a}{2}\left\langle \hat{\phi}_+(t)\right\rangle - b\left\langle \hat{\phi}_z(t)\right\rangle \tag{14}$$

and

$$\frac{d}{dt}\left\langle \hat{\phi}_-(t)\right\rangle = -\frac{a}{2}\left\langle \hat{\phi}_-(t)\right\rangle + b\left\langle \hat{\phi}_z(t)\right\rangle. \tag{15}$$

Where $a = 2\gamma\left(N-M+\frac{1}{2}\right)$ and $b = \gamma M$. To find the steady state solution for the two-time correlated function in Equation (11), let's add the Equations (14) and (15) first and then we get the result as

$$\frac{d}{dt}\left(\left\langle \hat{\phi}_+(t)\right\rangle + \left\langle \hat{\phi}_-(t)\right\rangle\right) = -\frac{a}{2}\left(\left\langle \hat{\phi}_+(t)\right\rangle + \left\langle \hat{\phi}_-(t)\right\rangle\right). \tag{16}$$

Using separation of variable to write Equation (16) as

$$\int_t^{t+\tau}\frac{d\left(\left\langle \hat{\phi}_+(t)\right\rangle + \left\langle \hat{\phi}_-(t)\right\rangle\right)}{\left\langle \hat{\phi}_+(t)\right\rangle + \left\langle \hat{\phi}_-(t)\right\rangle} = -\frac{a}{2}\int_t^{t+\tau}dt. \tag{17}$$

Integrating Equation (17), both sides we get

$$\ln\left(\frac{\left\langle \hat{\phi}_+(t+\tau)\right\rangle + \left\langle \hat{\phi}_-(t+\tau)\right\rangle}{\left\langle \hat{\phi}_+(t)\right\rangle + \left\langle \hat{\phi}_-(t)\right\rangle}\right) = -\frac{a}{2}\tau. \tag{18}$$

Equation (19) can be rewritten in the form

$$\left\langle \hat{\phi}_+(t+\tau)\right\rangle + \left\langle \hat{\phi}_-(t+\tau)\right\rangle = \left[\left\langle \hat{\phi}_+(t)\right\rangle + \left\langle \hat{\phi}_-(t)\right\rangle\right]e^{-\frac{a}{2}\tau}. \tag{19}$$

If we multiply Equation (19) both sides by the operator $\hat{\phi}_+(t)$ and taking the assumption that at any time $t$ the atomic variable combinations $\left(\left\langle \hat{\phi}_+(t)\hat{\phi}_+(t)\right\rangle = \left\langle \hat{\phi}_-(t)\hat{\phi}_-(t)\right\rangle = 0\right)$ be vanish, we will have the following expression

$$\left\langle \hat{\phi}_+(t)\hat{\phi}_+(t+\tau)\right\rangle + \left\langle \hat{\phi}_+(t)\hat{\phi}_-(t+\tau)\right\rangle = \left\langle \hat{\phi}_+(t)\hat{\phi}_-(t)\right\rangle e^{-\frac{a}{2}\tau}. \tag{20}$$

According to the quantum regression theorem for two quantum operators $\hat{A}$ and $\hat{B}$, the two-time correlation is written as [11]

$$\langle \hat{A}(t)\hat{B}(t+\tau)\rangle = \langle \hat{A}(t)\hat{B}(t)\rangle G(\tau). \tag{21}$$

Employing this equation into Equation (20), one gets

$$\langle \hat{\phi}_+(t)\hat{\phi}_-(t+\tau)\rangle = \langle \hat{\phi}_+(t)\hat{\phi}_-(t)\rangle e^{-\frac{a}{2}\tau}. \tag{22}$$

Furthermore, using Equations (5), (6), (7) and (8) along with the cyclic property of trace, it is easy to verify for a tow-level atom interaction with squeezed vacuum reservoir that

$$\frac{d}{dt}\rho_a = -\gamma(2N+1)\rho_a + \gamma N. \tag{23}$$

In this Equation $\gamma N$ is the rate of stimulated emission whereas $\gamma$ represents spontaneous emission rate. At steady state, the solution of Equation (23) is

$$\rho_a(\infty) = \frac{N}{2N+1}. \tag{24}$$

In view of Equations (23) and (24), the power spectrum described in Equation (11) can be put in the form of

$$P(\omega) = \frac{2N}{2N+1} \operatorname{Re}\int_0^\infty d\tau e^{-(a/2 - i(\omega-\omega_0))\tau}, \tag{25}$$

so that carrying out the integration, one readily finds

$$P(\omega) = \left(\frac{1}{2N+1}\right)\left(\frac{2N}{(\omega-\omega_0)^2 + \gamma^2(N-M+\tfrac{1}{2})^2}\right). \tag{26}$$

This equation describes the power spectrum of the fluorescent light emitted by a two-level atom interact with squeezed vacuum reservoir. If we plot the spectrum we will generate the graph depicted in Figure 1.

## 4. Results and discussions

From the figure 1a, we note that the power spectrum of a fluorescent light emitted from a two-level atom has a peak point at the frequency $\omega - \omega_0$. On the other hand, we see that the sharpness of the spectrum decreases as we go far from the point at which the transition frequency equals to the frequency of a light mode. This must be due to the decrement of the correlation between the two-time correlations. Moreover, the graph indicates that the mean number of photon in the reservoir mode has an effect on the sharpness of the power spectrum. One effect of the squeezed vacuum reservoir is to deform the sharpness of the spectrum. This is explained as when the mean photon number from the squeezed vacuum reservoir increases the power spectrum of the light emitted by a two-level atom deforms. But it is important to see from Equation (26) that the power spectrum is highly depending on the mean photon number of squeezed vacuum reservoir. The existence of these photons in the reservoir is very important for stimulated emission by the atom. This situation takes place when the atom is initially in the upper electronic energy state and the photon from the reservoir hits the atom which has resonant frequency with atomic transition frequency. Here we have done the mathematical equation behind such conditions which has to be true for the selected model.

If the reservoir is thermal, it is possible to generating the expression of the power spectrum for fluorescent light by simply setting $M = 0$ and $N = \bar{n}$ where $\bar{n}$ is the mean photon number for thermal reservoir. Thus we have

$$P(\omega) = \left(\frac{1}{2\bar{n}+1}\right)\left(\frac{2\bar{n}}{(\omega-\omega_0)^2 + \gamma^2(\bar{n}+\tfrac{1}{2})^2}\right). \tag{27}$$

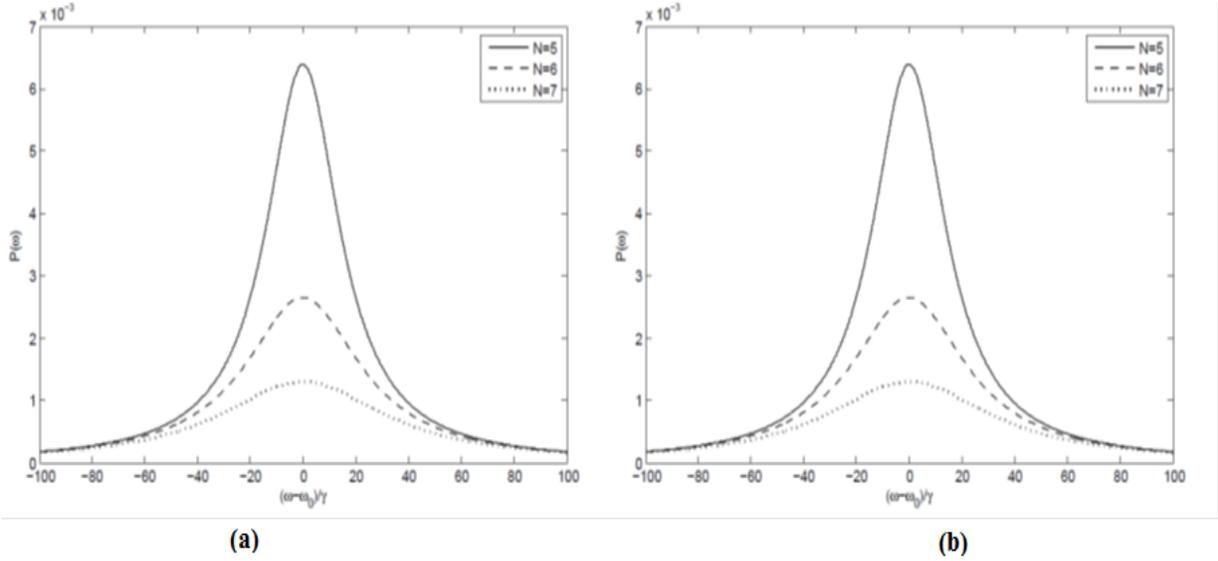

(a)  (b)

Figure 1. Plots in (a) the power spectrum of the light emitted by a two-level atom interacting with a squeezed vacuum reservoir for $N=5$ (solid curve), $N=6$ (dashed curve) and $N=7$ (dotted curve); Plots in (b) the power spectrum of the light emitted by a two-level atom interacting with a thermal reservoir for $\bar{n}=5$ (solid curve), $\bar{n}=6$ (dashed curve) and $\bar{n}=7$ (dotted curve).

As we observe the scale of the power spectrum in Figure 1(a) is in the order of $10^{-3}$ while Figure 1(b) is in the order of $10^{-2}$ for the same parameters used in both figures. Thus, the power spectrum a fluorescent light emitted by the atom coupled to thermal reservoir is greater than that of squeezed vacuum reservoir.

## 5. Conclusions

The paper addresses, the dynamics of a power spectrum emitted by a two-level atom interacts with the reservoir sub-modes with the aid of evolution of atomic variables. By driving the master equation that governs the system under consideration, we have obtained the time evolution for the atomic variable. Finally, we have got the two-time correlation for the raising and lowering atomic operators which is the core problem of this paper and using the result (Equation (22)) power spectrum for a fluorescent light is obtained. It is identified that the presence of reservoir mode is responsible to stimulating emission of photon from an atom. Moreover, the power spectrum of a fluorescent light is greater when the atom interacts with thermal reservoir than squeezed vacuum reservoir. The other outstanding point in this paper is that the power spectrum of a fluorescent light is turn out to be a single peak whether the reservoir is squeezed vacuum or thermal. The result shows only there will be a single peak at the point at which the frequency of the light is equals to the central Lorentz's frequency [1]. This must be due to high correlation is induced between the rising and lowering atomic operators expressed in Equation

### Acknowledgments

We thank Getnet Melese (Assistant Professor of physics at Debre Tabor University) for technical support in design of the study; in the collection, analyses, or interpretation of data; in the writing of the manuscript.